\documentclass[lettersize,journal]{IEEEtran}
\usepackage{amsmath,amsfonts}
\usepackage{algorithmic}
\usepackage{algorithm}
\usepackage{array}
\usepackage[caption=false,font=normalsize,labelfont=sf,textfont=sf]{subfig}
\usepackage{textcomp}
\usepackage{stfloats}
\usepackage{url}
\usepackage{verbatim}
\usepackage{graphicx}
\usepackage{cite}
\begin{document}

\title{Beam-Shape Effects and Noise Removal from \\ THz Time-Domain Images in Reflection Geometry in the 0.25 - 6 THz Range} 

\author{Marina Ljubenovi\'c,
        Alessia Artesani,
        Stefano Bonetti, and
        Arianna Traviglia

\thanks{This work has been submitted to the IEEE for possible publication. Copyright may be transferred without notice, after which this version may no longer be accessible.}
\thanks{This project has received funding from the European Union's Horizon 2020 research and innovation programme under grant agreement No. 101026453.}
\thanks{Marina Ljubenovi\'c, Alessia Artesani, and Arianna Traviglia are with the Center for Cultural Heritage Technology, Istituto Italiano di Tecnologia, Venice, Italy.}
\thanks{Stefano Bonetti is with the Department of Molecular Sciences and Nanosystems, Ca'Foscari University of Venice, Venice, Italy and the Department of Physics, Stockholm University, Stockholm, Sweden.}
}



\maketitle

\begin{abstract}
The increasing need of restoring high-resolution Hyper-Spectral (HS) images is determining a growing reliance on Computer Vision-based processing to enhance the clarity of the image content. HS images can, in fact, suffer from degradation effects or artifacts caused by instrument limitations. This paper focuses on a procedure aimed at reducing the degradation effects, frequency-dependent blur and noise, in Terahertz Time-Domain Spectroscopy (THz-TDS) images in reflection geometry. It describes the application of a joint deblurring and denoising approach that had been previously proved to be effective for the restoration of THz-TDS images in transmission geometry, but that had never been tested in reflection modality. This mode is often the only one that can be effectively used in most cases, for example when analyzing objects that are either opaque in the THz range, or that cannot be displaced from their location (e.g., museums), such as those of cultural interest.  Compared to transmission mode, reflection geometry introduces, however, further distortion to THz data, neglected in existing literature.  In this work, we successfully implement image deblurring and denoising of both uniform-shape samples (a contemporary 1 Euro cent coin and an inlaid pendant) and samples with the uneven reliefs and corrosion products on the surface which make the analysis of the object particularly complex (an ancient Roman silver coin). The study demonstrates the ability of image processing to restore data in the 0.25 - 6 THz range, spanning over more than four octaves, and providing the foundation for future analytical approaches of cultural heritage using the far-infrared spectrum still not sufficiently investigated in literature. 
\end{abstract}


\section{Introduction}
\label{sec:intro}

 

\IEEEPARstart{S}{pectroscopy} and imaging in the terahertz (THz) spectral range (typically 0.1 to 10 THz) is gaining attention in many disciplines such as biomedicine, agriculture, security, and communication services \cite{2011_Wietzke_Structural,2018_Neu_Tutorial}.
One of the major applications of THz spectroscopy and imaging is in noninvasive detection of material composition and retrieval of complex surface topography of objects \cite{Fukunaga2011,Filippidis2012,Manceau2008}. 
A THz signal exhibits many unique properties: i) its low-photon energy facilitates nondestructive and non-ionizing testing \cite{2008_Nazarov_Biology}, ii) it has the ability to penetrate nondielectic materials (e.g., paper, plastic, etc.); iii) it can be used in reflection geometry for studying objects that are too thick for signal transmission, or highly reflecting (e.g., objects made of metal).


Terahertz Time-Domain Spectroscopy (THz-TDS) employs short pulses of THz radiation to probe dielectric response of materials in the far-infrared region and reconstruct objects' inner structure, giving as result a 3-dimensional hyperspectral (HS) data-cube, providing information on both surface and inner structures.  
Despite the fact that THz-TDS imaging offers a unique means to probe materials' properties, it poses also major challenges in terms of data processing because of the multiple sources of image degradation that make difficult the analysis of the THz signals \cite{2020_Ljubenovic_Joint}.
THz-TDS images are in fact corrupted by frequency-dependent beam-shape induced blur and noise with additional signal distortions introduced by scattering losses (more information on degradation effects will be provided in Section \ref{sec:degradation}). All these image degradation effects remarkably reduce the frequency region of the resolved images and limit the recovering of sharp THz images. 

Some conventional image restoration algorithms can singularly tackle blur effects, noise, or low resolution, thus overcoming only partially physical limitations determined by the THz-TDS imaging technique.  
The joint removal of several degradation effects in THz HS images represents still a challenge due to several reasons: a frequency dependency of blur and noise; the presence of blocking artefacts introduced by the step size of a scanning system; and existence of additional effects such as reflection and refraction \cite{2020_Ljubenovic_Joint}.
At present, there is a limited number of efficient computational strategies to address some of the THz imaging problems and the majority of the proposed approaches have dealt with single-frequency measurements, such as THz images in continuous wave mode or volumetric reconstruction with THz computed tomography. 

THz-TDS imaging can be carried in transmission and reflection geometries. In transmission geometry, the temporal profile of the electric field transmitted through the investigated sample is measured and compared with a reference spectrum obtained from the free path. However, many samples possess significantly high absorption properties in the THz range or are highly reflective, situations that limit or completely impede the use of THz imaging in transmission geometry. Furthermore, for samples with a complex structure, absorption of the radiation in the high frequency range limits the bandwidth to 2-2.5 THz in most cases. All these issues are minimized or completely removed performing the measurements in reflection geometry, which allows to measure virtually any sample, and with a much larger bandwidth. Furthermore, it has been demonstrated that reflection imaging is more sensitive to sample adhesion (e.g., deviation in thickness of paint and coating) that can cause variation in the phase of the THz signal [8]. However, this greater sensitivity is also a potential source of new degradation artefacts in the resulting image, such as discontinuous spots. Other degradation effects that could be present in reflection geometry, and are not present in the transmission one, includes distortion and signal losses due to refraction, diffraction, edge effects, specular reflection and Fabry-Perot effects, which might heavily corrupt the THz images. It follows that in order to fully exploit the great potential of THz imaging in reflection geometry, specific corrections of these degradation effects need to be implemented.

This work describes the application to THz-TDS images of an algorithm that has been recently developed and successfully applied for joint THz image deblurring and denoising \cite{2020_Ljubenovic_Joint}. 
The algorithm was originally developed for THz imaging in the transmission mode and validated on objects having a simple shape (e.g., leaf, ring, plate). Here, it is adapted for the first time for THz reflection images and tested on samples with more complex material structures and textures. 
Section \ref{sec:CV_SotA} starts by introducing existing image processing methods that are developed or applied to THz data and their limitations. The experimental setup is introduced in Section \ref{sec:acq}, followed by the introduction of degradation effects in Section \ref{sec:degradation}.
The efficiency of the joint deblurring and denoising approach is tested on highly reflective objects (Section \ref{sec:THz_processing}), and then applied to the study of an ancient coin (Section \ref{sec:thz_in_ch_res}). 
The joint deblurring and denoising approach, tailored to frequency- dependent blur degradation, utilizes the high correlation of spectral bands to increase their availability for the further analysis. 

\section{Image Processing Methods for THz Data}
\label{sec:CV_SotA}
Initial attempts of THz image restoration have employed conventional image restoration methods originally developed for other imaging modalities (e.g., X-ray, RGB images)  \cite{2014_Xu_High-Resolution}. These include but are not limited to iterative backprojection, Richardson-Lucy method  \cite{1972_Richardson_Bayesian},\cite{1974_Lucy_Iterative}, and 2D wavelet decomposition reconstruction \cite{2009_Mallat_Wavelet}. For instance, Li \textit{et al.} in \cite{2008_Li_SuperResolution} applied a well-known Richardson-Lucy deblurring algorithm to remove beam shape effects from a single-frequency THz images improving the boundary sharpness of binary samples, but introducing artefacts in the presence of fine details. 
Those conventional deblurring algorithms have been designed to improve resolution of images that follow different statistics: they are demonstrated to be inadequate for THz data mostly due to the huge difference in the assumed prior knowledge imposed on a sharp image and the degradation effects (e.g., blur and noise).

At first, the development of procedures aiming at improving THz data was mainly focused on specific applications such as THz-based computed tomography (THz-CT) or subwavelength microscopy. THz-CT data often require sophisticated algorithms to manage volume reconstruction from a projection domain. One of the first methods tailored to THz-CT was introduced by Recur \textit{et al.} in \cite{2012_Recur_CT}. The method combines the CT reconstruction approaches (e.g., back-projection of filtered projections, simultaneous algebraic reconstruction technique, and ordered subsets expectation maximisation) with a convolution filter to mimic the THz beam effects. The influence of these effects can be further removed by employing the Wiener deconvolution approach  \cite{1985_Dhawan_Wiener,2010_Popescu_Point}. Some attempts were made for increasing the resolution of images with a multilayered structure by applying deconvolution in the time domain \cite{2009_Takayanagi_HR_Tomography}. 
These methods can tackle noise-free single-frequency THz images or signals, but do not take into account the differences imposed by the dependence of the noise and blur effects on the frequency value. These methods are not suitable for a broadband THz imaging systems and the complex structure of its beam as they are considering images corresponding to a single THz frequency as independent. 
To beat the diffraction limit in subwavelength microscopy, a super-resolution technique for THz images was proposed in  \cite{2020_Guerboukha_SR_Microscopy}. This technique utilizes specially designed artificial fluorophores in the form of optimal mask sets and stochastic fluctuations in the intensity of the fluorophores limiting its broad usage.

The complex structure of a broadband THz beam and its effects was studied in \cite{2016_Kiarash_THz_Equation}. The authors modelled a 3D THz point spread function (PSF) and THz imaging equation, further utilized to enhance single-frequency THz images by applying a simple deconvolution approach (i.e., Matlab in-built function, deconv). Utilisation of a more advanced deconvolution procedure for tackling blur degradation in frequency modulated continuous wave THz images was introduced by Wong \textit{et al.} in \cite{2019_Wong_Computational}. This paper introduced two steps leading to THz image enhancement: high-precision depth estimation followed by 2D deconvolution with off-the-shelf methods \cite{2013_Xu_Unnatural}.
Although these approaches worked reasonably well for noise-free images and simple-shape samples, due to the ill-conditioned nature of the convolution, they introduced severe error in a presence of even small amount of noise.

Deep neural networks (DNNs) have also been used to ameliorate THz data using robust restoration approaches, but rarely considering the frequency-dependence of the THz-TDS images during the training process.  They were mostly used for 2D THz image deblurring  \cite{2020_Ljubenovic_THzCNN} and super-resolution \cite{2017_Li_THzSuperRes,2019_Long_THzSuperRes} or for specific applications, such as biological product analysis \cite{2021_Lei_TimeDomain_SR}. The main limitation of DNN-based methods is the need for a large number of THz images (hundreds or thousands) for accurately training the network: unavailability of a sufficient number of THz images has, as a consequence, limited their application. 


One of the domains in which the removal of blur and noise degradation of HS images have been significantly developed is that of airborne Remote Sensing imagery: here, the clarity of the HS images is key for their interpretation and the quality of the spectral datum pivotal to further image processing. This has led to the development of several methods to improve the quality of the image, such as denoising (HySime \cite{2007_Nascimento_HySime} and FastHyDe \cite{2018_Zhuang_FastHyDe}) and deblurring methods \cite{2013_Liao_HSIDeblurring} that leverage some characteristics of the HS images like low-rank and self-similarity of image patches. 
In particular, FastHyDe demonstrated to be especially successful in denoising HS images affected by types of noise such as Gaussian independent and identically distributed (i.i.d.), Gaussian non-i.i.d., and Poisson noise. Notwithstanding their potential transferability in other domains, these image processing approaches have not found so far application in close-range THz imaging. 

An approach that considers bands of THz HS images jointly has been introduced in \cite{2020_Ljubenovic_Joint}. Similar to FastHyDe, the method is based on the low-rank and self-similarity properties of HS images and it includes two additional expansions that make it suitable for THz data: i) a specialized THz beam modelling process and ii) a deblurring step based on well-known, robust deblurring approaches.
The joint deblurring and denoising approach represents an effective measure to counter frequency-dependent blur and three different noise types present in time-domain THz images acquired in the transmission mode, but, to the authors' knowledge, it has never been applied in the reflection mode and on objects with irregular geometries. Our work focuses on demonstrating the application of the HS algorithms for THz imagery on uneven object samples, as an archaeological coin. The analysis on these kind of samples is complicated by the irregular surface, the asymmetry of their thickness, the presence of irregular reliefs on the samples, the variability of their material compositions (often unknown and/or multiform) and contribution of both reflecting and absorbing materials. \\
Our research has entailed the application of a joint denoising and deblurring approach to improve the quality of THz images in reflection geometry: the paper critically discusses the acquisition procedure, the image restoration requirements, and the limitations of the proposed framework. The degradation effects induced by the THz beam size are evaluated in order to determine the point-spread function (PSF) of the THz-TDS system necessary to remove image degradation. 
Band-by-band deblurring, HS image denoising, and a method that remove blur and noise jointly are initially tested on objects characterized by uniform material composition and relatively unpatterned shape (a 1 Euro cent coin –€0.01– and an inlaid silver pendant).  After demonstrating that only the joint deblurring and denoising method achieves good results through all bands, and after evaluating its limitations and advantages, the method is directly applied to an ancient Roman silver coin and the differences between the two types of samples (i.e., contemporary and archaeological objects) are highlighted. 


\section{Experimental Setup and Acquisition Procedure}
\label{sec:acq}

\begin{figure*} [!t]
\centering
\includegraphics[width=0.9\textwidth]{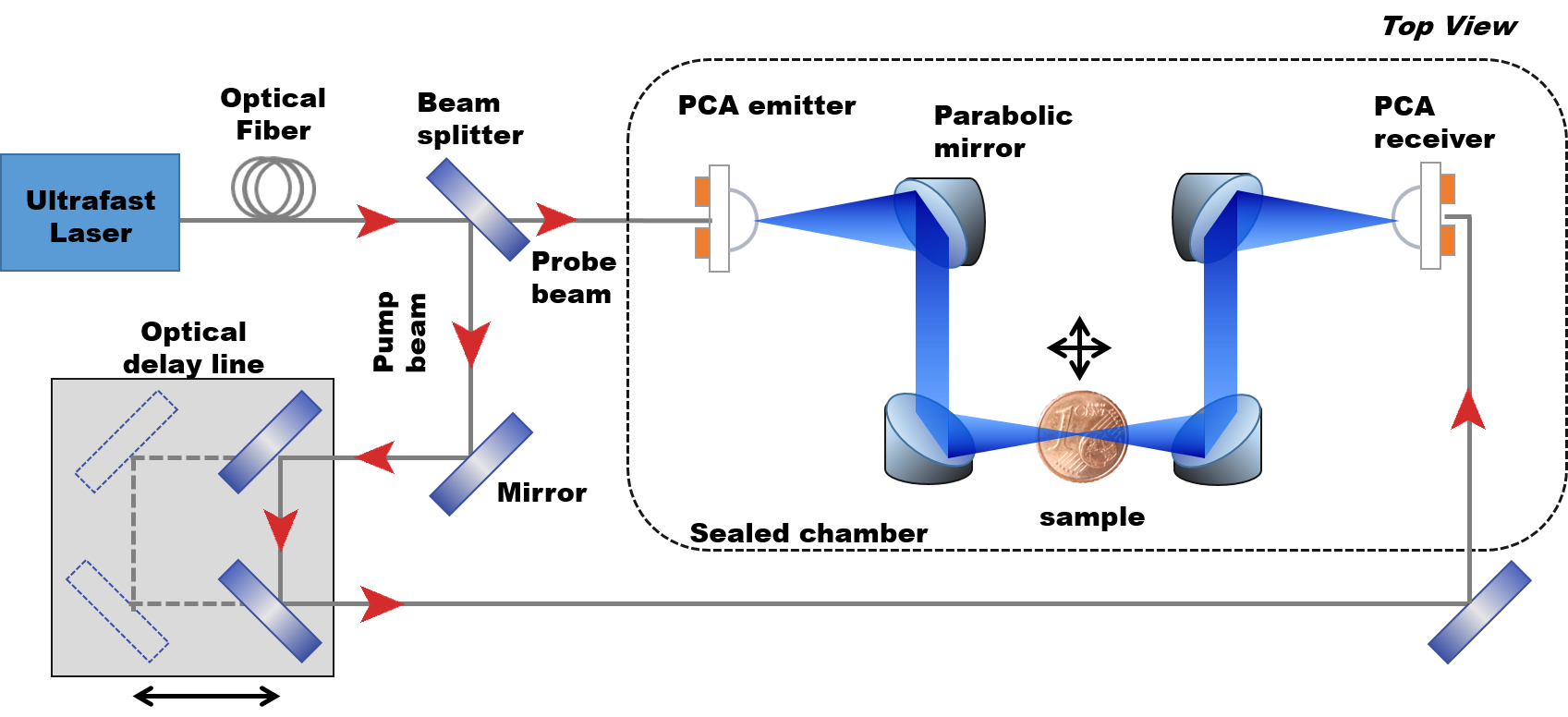}
\caption{Scheme of the THz-TDS system in the reflection mode. The instrument is based on an ultrafast laser source, emitter and receiver antennas and motorized raster-scanning stage, where the sample is positioned for measurements in reflection geometry.}
\label{fig:thz_tds_system}       
\end{figure*}

This work has been completed using a commercial THz-TDS system  (TOPTICA TeraFlash Pro, Fig. \ref{fig:thz_tds_system}). 
The system is based on a mode-locking Erbium-doped fiber laser ($\lambda$ = 1560 $\pm$  10 nm), with pulse width of 50-60 fs and repetition rate of 100 MHz (TOPTICA FemtoFErb THz FD6.5). The ultrafast laser pulse is delivered via a single-mode fiber onto an InGaAs/InAlAs-based photoconductive antenna that in turn produces an electromagnetic pulse with the frequency spectrum in the THz range. The system is capable of generating a spectrum width that spans 6 THz, has the highest power being emitted at frequencies around 1 THz. The precise reconstruction of the time axis and the reduced jitter noise of the system enables to achieve a good dynamic range below 3.5 THz (in nitrogen atmosphere), with a peak dynamic range of more than 90 dB. The measured THz electric field is Fourier transformed to yield the phase and the amplitude of the THz pulse.


The system relies on point-like measurements. The analysis of a surface is obtained in the reflection mode by raster-scanning the sample placed in the focal plane of the two parabolic mirrors, using a motorized XY stage that is synchronized  with the THz excitation.  The image is obtained by continuous line scanning along the X direction (from left to right) with a step size of $\Delta$x, and by step line along the Y direction of $\Delta$y. For the TOPTICA TeraFlash Pro, the step sizes in both directions can be fixed equal to 0.1, 0.2, 0.5, 1 or 2 mm. Whereas the frequency resolution of the THz‐TDS is given by the time step and the number of data points in the time‐domain measurement, the spatial resolution is limited by the point spread function (PSF) of the THz beam (as discussed in Section \ref{sec:degradation}), and also by the acquisition step size along the X and Y axis.

\begin{figure}[!t]
\centering
  \includegraphics[width=0.48\textwidth]{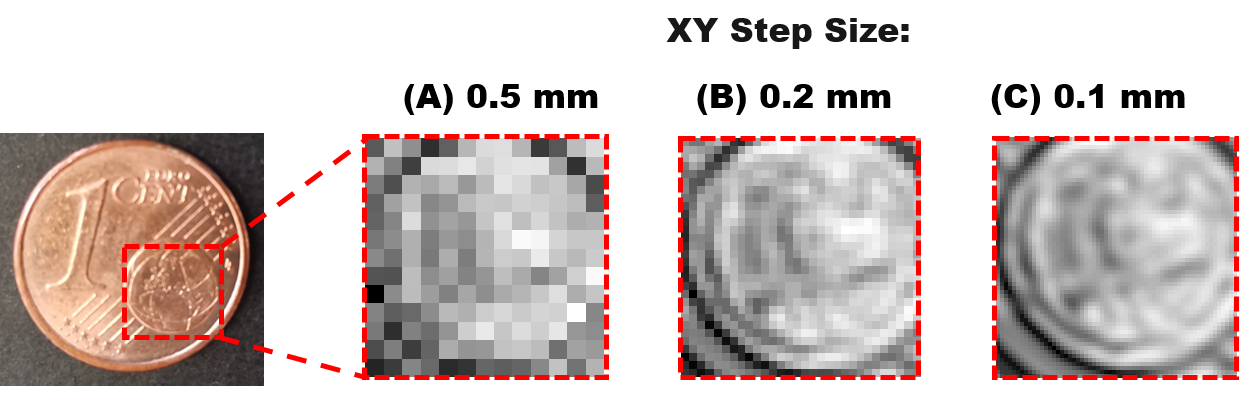}
\caption{Example of the amplitude signal $A(\omega)$ of a raster-scanned sample taken with three different $\Delta$x and $\Delta$y step sizes, equal to (A) 0.5 mm, (B) 0.2 mm, (C) 0.1 mm. The frequency is fixed at 2.72 THz.}
\label{fig:step_sizes}       
\end{figure}
In order to show the difference between the step size choice, a portion of a 1 Euro cent coin has been raster-scanned with step size set equal to 0.5 (A), 0.2 (B), and 0.1 (C) mm, respectively (Fig.\ref{fig:step_sizes}). With the biggest step size, the detector loses information leading to severe blocking degradation in the resulting image. Due to this degradation, smoothness of the object edges is lost. With the smallest step size, the THz images reaches a higher level of details, but the acquisition procedure lasts significantly longer (e.g., 3 min (B) vs. 32 min (C) for scanning $5 \times 5$ mm) and a data size increases ($\sim 12$ MB (B) vs. $\sim$ 120 MB (C)), while missing in improving significantly the spatial resolution due to the beam shape effects. For the considered sample, the optimum between the choice of the step size and the acquisition time is the option (B), whereas the value of the step size should be carefully considered taking into account aspects such as dimension of the samples and its features, and the desired application results. 


\section{Degradation Effects}
\label{sec:degradation}

\begin{figure*}[!t]
\centering
  \includegraphics[width=1\textwidth]{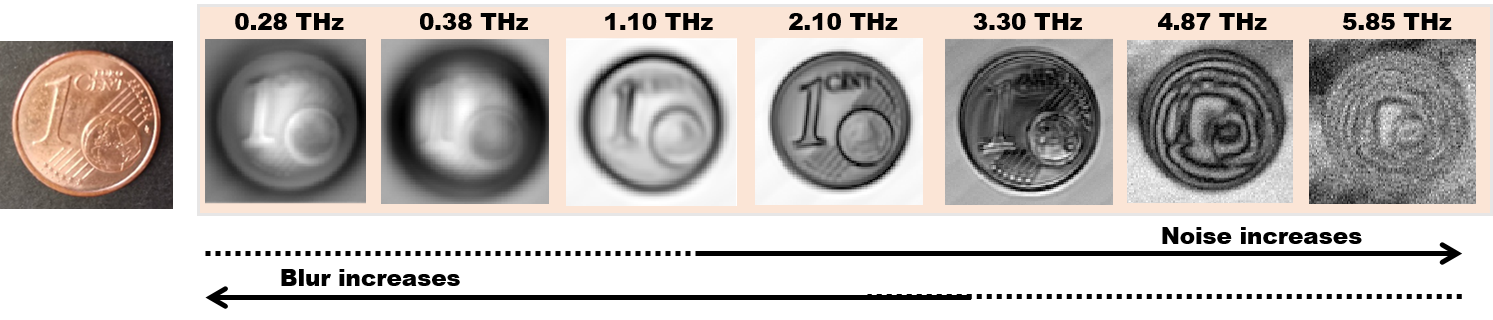}
\caption{Typical degradation effects in the representation of the amplitude $A(\omega)$ images. The images are corrupted by noise that increases with the frequency, while at lower frequency value, the blur effects are more dominant.The frequency region with relatively low noise and blur degradation effects stays between 0.5 and 3.5 THz.}
\label{fig:raw_data}       
\end{figure*}

The THz system design, with focusing optics and beam forming process, leads to frequency-dependent degradation effects that corrupt the amplitude images. The main causes for the degradation effects that can alter the spatial information of the THz images are mainly: i) the THz beam-shape; ii) the system-dependent noise; iii) losses due to the propagation of the THz wave through the sample (caused by reflection and/or refraction); iv) the scattering and absorption properties of the sample. Whereas the latter two degradation effects are intrinsically dependent on the sample and hard to model, the consequences of the THz beam-shape and the noise of the system can be estimated by introducing simple approximations. Thus, the restoration algorithms are based on the following assumptions for the blur and noise effects. \\

\textbf{Blur effects.} The intensity profile of the THz beam is assumed to have a Gaussian-shape \cite{2012_Recur_CT}. Under this approximation, the transverse profile of the THz intensity is described by
\begin{equation}
\label{equ_intensity}
I(r,z) = \frac{P}{\pi w^2(z)/2}\text{exp} \left(-2\frac{r^2}{w^2(z)}\right),
\end{equation}
where $P$ is the beam power, $w(z)$ is the beam radius, $z$ is the propagating direction and $r$ is the distance from the beam axis. The radius of the beam varies along the propagating direction as
\begin{equation}
\label{equ_radius}
w(z) = w_0 \sqrt{1 + \left(\frac{z}{z_R}\right)^2},
\end{equation}
where  $w_0$ is minimum beam radius (half of the beam waist) and $z_R = \pi w_0^2/\lambda$ represents the Rayleigh length at wavelength $\lambda$ that determines the length over which the beam can propagate without diverging significantly. In Gaussian beam approximation, the beam waist $2w_0$  depends on the wavelength $\lambda$ accordingly to the \cite{1998_Goldsmith_GaussianBeam}:
\begin{equation}
\label{equ:relation_w0_lambda}
2w_0 = \frac{4}{\pi} \lambda \frac{f_L}{D},
\end{equation}
where $f_L$ and $D$ represent a focal length and a diameter of the focusing THz lens. The size of the THz beam waist is strongly dependent on the frequency and the focusing optics of the THz system. The first degradation effect associated to the beam-shape consequently affects the resolution of the HS image and introduces blurring. In particular, every pixel in the image spatial domain is blurred following a 2D Gaussian PSF corresponding to an intersection of the THz beam with an orthogonal plane (represented by the surface of the sample). As the THz beam waist is wider at lower frequencies, the blurring effects will be more important in the lower region of the THz frequency range, as illustrated in Fig. \ref{fig:raw_data}. On the contrary, at higher frequencies, a THz beam waist is smaller and the resulting images are sharper. \\

\textbf{Noise effects.} The principal degradation effect at high frequency is the noise determined by the instrument. The signal from a THz-TDS system is corrupted by, at least, three noise components \cite{1990_vanExter} \cite{2000_Duvillaret_Noise}. The first noise component is caused by the transmitting antenna, which represents the principal noise source. The second contribution is Poisson noise in the detector, which usually originates from the discrete nature of the electric charge. The third component summons other signal-independent noise in the detector such as amplification noise, laser noise, thermal noise, etc. 
All these noise components are severely decreasing the signal-to-noise ratio with the increase of frequency, making the images corresponding to higher frequency bands impractical for further analysis. For the instrument employed, this occurs for frequencies above 4.5 THz (see Fig. \ref{fig:raw_data}).\\

\begin{figure}[!t]
    \centering
    \includegraphics[width=0.47\textwidth]{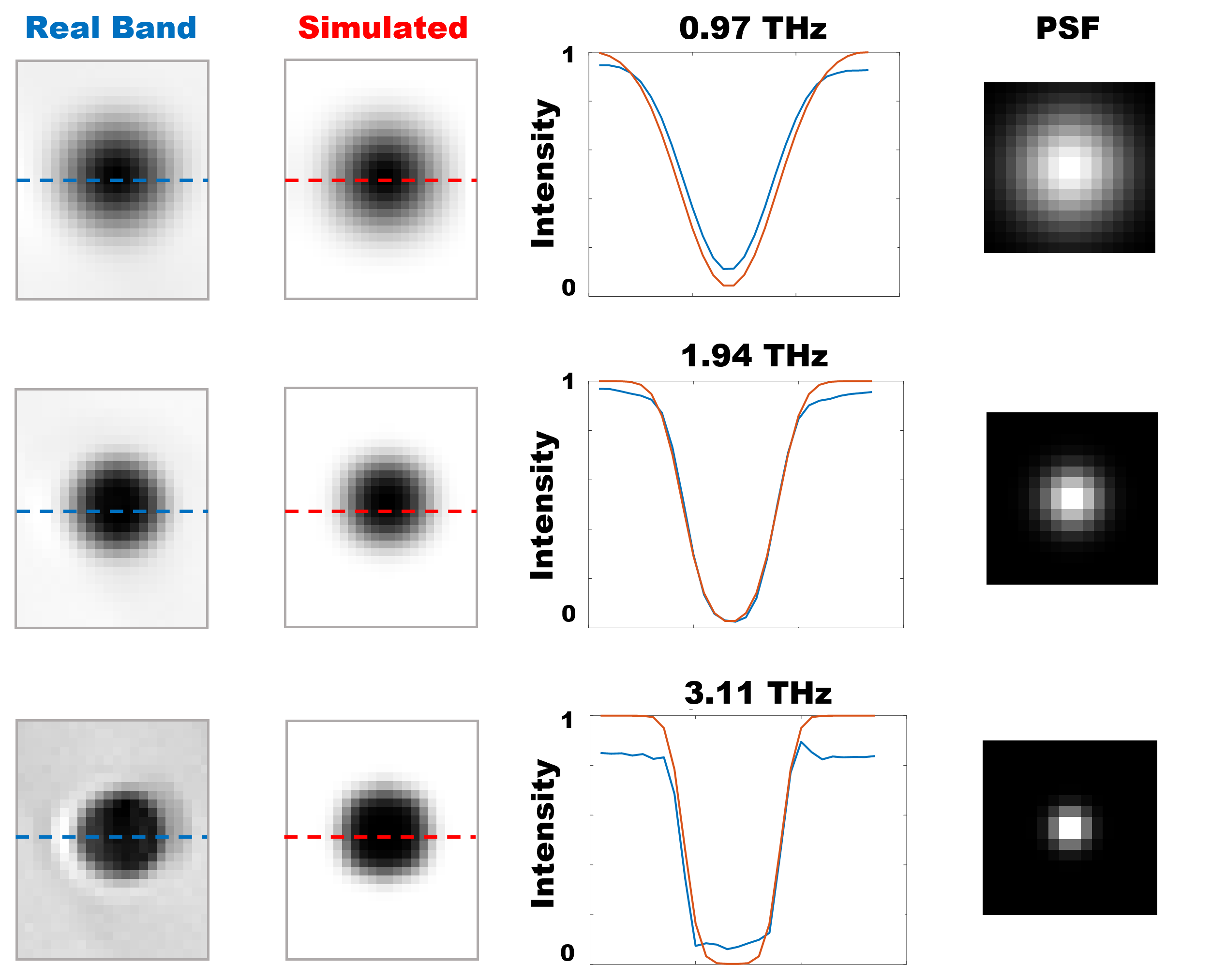}
    \caption{Comparison between simulated reference images and experimental measurements of a hole on a metallic plate at three THz frequencies. From left to right: the denoised bands of raw data on three different frequencies (0.97, 1.94, and 3.11 THz); simulated bands corresponding to the same frequencies; intensity vectors corresponding to the middle cross-section of real (blue) and simulated (red) bands; and theoretical PSFs.}
    \label{fig:blurring filter}       
\end{figure}

In order to validate the above-mentioned Gaussian modelling assumption implemented in the algorithm for working in reflection geometry, synthetic reference images of a circle blurred by a Gaussian model have been compared with the real bands of THz time-domain images of a hole on a metallic plate. The theoretical THz beam waist, calculated considering that the instrument has two focusing mirrors with 1-inch diameter and 4-inches length, is $w_0 \approx 2.547 \lambda$. The experiments were performed considering step size $\Delta x$ and $\Delta y$ equal to 0.2 mm, and beam diameter equals to the beam waist. The results are presented in Fig. \ref{fig:blurring filter}. 
On the lower and medium frequencies (i.e., 0.97 and 1.94 THz), the difference in intensity between synthetic and real bands is mostly visible at the borders of the circle, while on the higher frequency (i.e., 3.11 THz) the small difference is additionally visible in the center, this being most likely due to the influence of noise. Thus, the performance of the overall method is mostly reduced near the edges. The test shows the theoretical PSFs (intersection of an orthogonal plane with the beam calculated by applying the equations (\ref{equ_intensity}) and (\ref{equ_radius})) are comparable with the real blurring degradation.

\section{Image Restoration for THz Images}
\label{sec:THz_processing}

The goal of this work is restoring the amplitude images, which have the form of a hypercube with the third dimension corresponding to the THz frequencies. These amplitude images are modelled as: 
\begin{equation}
\label{equ_obs_model_H}
\textbf{Y} = \textbf{X} \textbf{H} + \textbf{N},
\end{equation}
where $\textbf{Y} \in \mathbb{R}^{b \times n}$ represents the degraded HS amplitude image with the rows containing $b$ spectral bands. Every band is an image with $n$ pixels corresponding to the amplitude intensity. $\textbf{X} \in \mathbb{R}^{b \times n}$  represents an underlying clean HS image (without noise and blur effects), while $\textbf{H} \in \mathbb{R}^{n \times n}$  is the blurring operator that models the convolution of a single band and the corresponding PSF. In the defined model, the same blurring operator is assumed over bands.  Finally, $\textbf{N} \in \mathbb{R}^{b \times n}$ represents additive white Gaussian i.i.d. noise. 

Prior to the application of the joint denoising and deblurring approach, the restoration of the degraded images is undertaken by conventional band-by-band deblurring (e.g., Richardson-Lucy or Wiener filtering) and denoising (FastHyDe) algorithms to demonstrate how the two algorithms only partially improve the image quality. The joint application of the denoising and deblurring approach is then described, taking into account the variable effects for each band. Noteworthy, the joint deblurring and denoising algorithm does not simple combine the two methods (i.e., FastHyDe and Richardson-Lucy), but follows a number of carefully designed steps, described in Subsection \ref{subsec:joint_processing}C.


\textbf{\subsection{Band-By-Band Deblurring Algorothm}}
\label{subsec:2d_deblurring}

A band-by-band deblurring is performed to remove blurring effects from THz-TDS images, where the deblurring problem for one band is modeled as
\begin{equation}
\label{equ_one_band_model}
\textbf{y}_i =  \textbf{H} \textbf{x}_i + \textbf{n}.
\end{equation}
Here, $\textbf{y}_i$, $\textbf{x}_i$, and $\textbf{n}$ represent an observed vectorized image of the $i$-th band ($i$-th row of $\textbf{Y}$), a sharp image of the $i$-th band, and noise respectively. 

Three conventional approaches are tested on the 1 cent coin and the silver pendant to perform band-by-band deblurring, specifically:  1) Richardson-Lucy deblurring \cite{1972_Richardson_Bayesian}, \cite{1974_Lucy_Iterative}; 2) Krishnan \textit{et al.}'s deblurring with a prior knowledge depicting heavy-tailed distribution of gradients in natural scenes (hyper-Laplacian prior) \cite{2009_Krishnan_Fast}; 3) Deblurring based on Wiener filtering \cite{1985_Dhawan_Wiener}. 
These three approaches are selected as the ones most commonly used in the literature for THz beam shape effects removal. 

\begin{figure*}[t!]
    \centering 
    \includegraphics[width=1\textwidth]{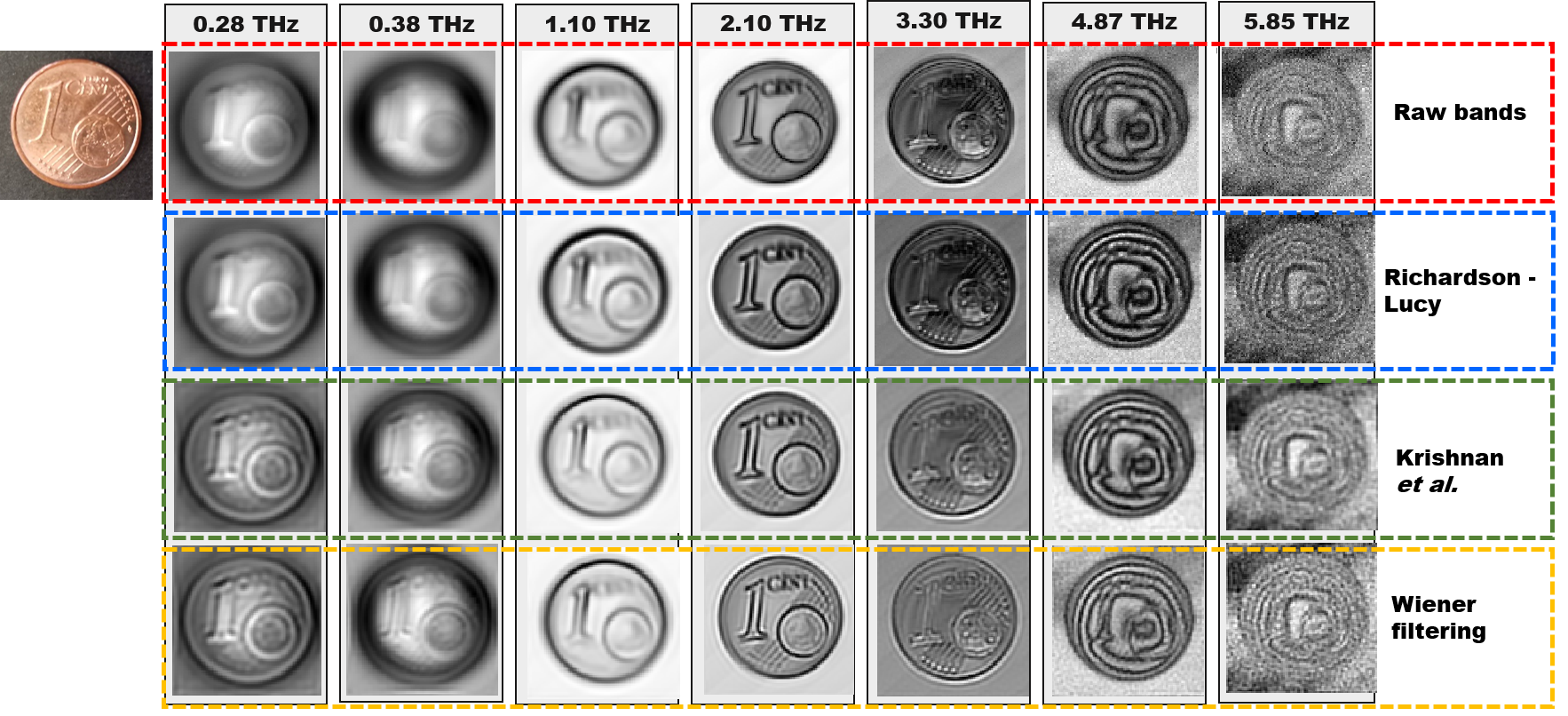} 
    \includegraphics[width=1\textwidth]{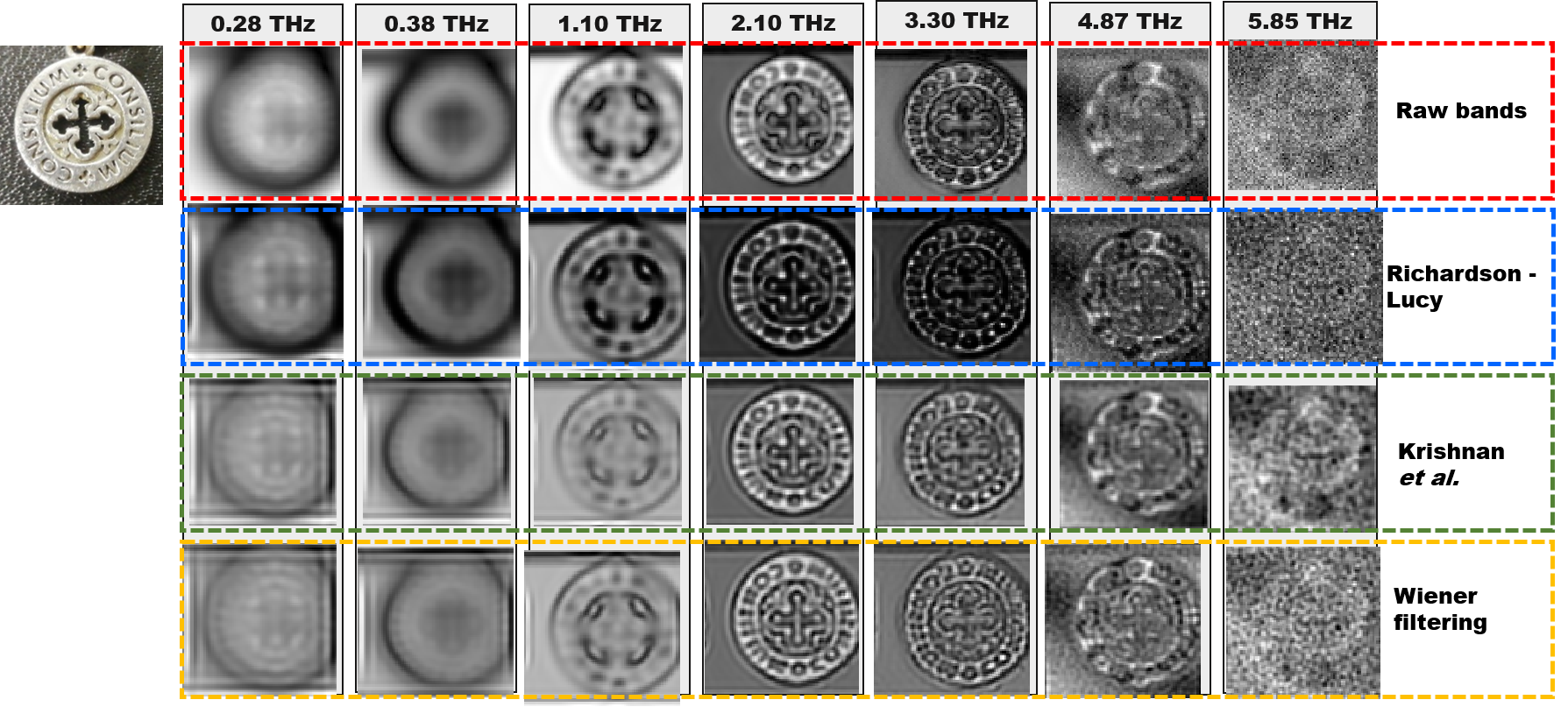} 
    \caption{Results of band-by-band deblurring approaches for restoring amplitude images on 1 Euro cent coin (top) and a silver pendant (bottom). The first rows corresponding to each sample show the optical image of the sample and selected row bands on seven different frequencies (red dashed line). The tested deconvolution approaches: 1) Richardson-Lucy iterative deconvolution \cite{1972_Richardson_Bayesian},\cite{1974_Lucy_Iterative} (blue dashed line); 2) Fast image deconvolution with a hyper-Laplacian priors by Krishnan \textit{et al.} \cite{2009_Krishnan_Fast} (green dashed line); 3) Image deblurring using Wiener filtering \cite{1964_Wiener_Extrapolation} (yellow dashed line).
    }
    \label{fig:2d_deblurring}      
\end{figure*}

The results obtained by three conventional deblurring approaches are presented in Figure \ref{fig:2d_deblurring} by displaying seven  selected bands of THz time-domain images corresponding to lower frequencies (0.28 and 0.38 THz), medium frequencies (1.1, 2.1, and 3.3 THz), and high frequencies (4.87 and 5.85 THz). 
The input parameters (the number of iterations, regularization parameters, etc.) are selected according to values reported in the original papers and additionally, when necessary, are adjusted to provide the best visual results.

The tested methods show some improvements for the selected medium frequencies and poor results when tested on bands corrupted by severe blur (the band corresponding to the low frequencies) or severe noise (the band corresponding to the high frequencies). The additional limitation when applying 2D-based deblurring approaches is hand-tuning of input parameters for every band separately (e.g., number of iterations for the Richardson-Lucy approach or a regularization parameter for the approach from Krishnan \textit{et al.}). This is because these band-by-band deblurring methods have been tailored to conventional RGB images and thus, in the original papers, values for input parameters are governed by the statistics of natural images, which are quite different than the one of THz images.
In addition, these methods usually require several input parameters that are different for different HS image bands and noise levels and therefore, in the case of THz HS images, challenging to properly set. 

\begin{figure*}[!t]
    \centering 
    \includegraphics[width=0.95\textwidth]{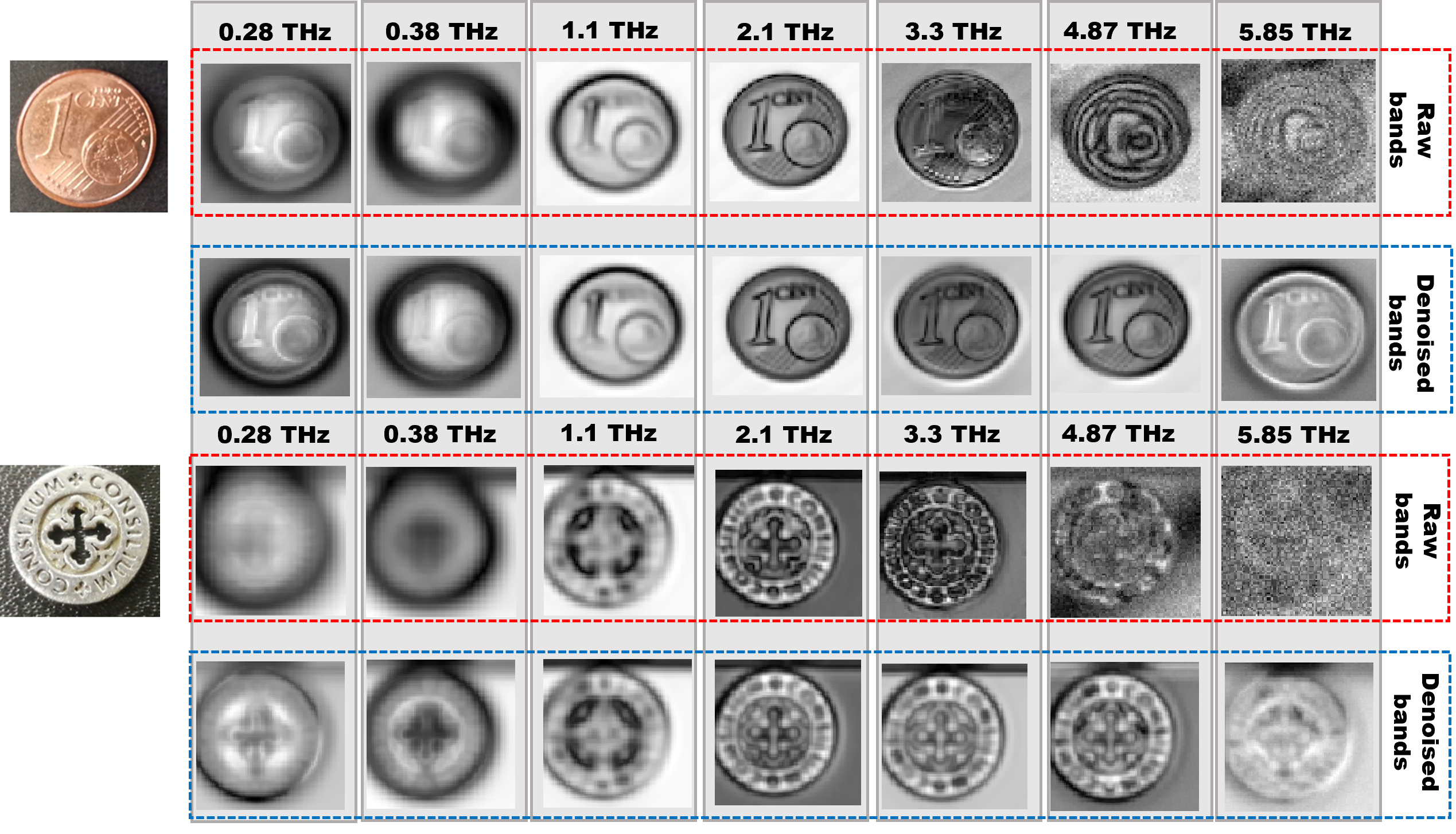} 
    \caption{Results of the FastHyDe method for restoring amplitude images on 1 Euro cent coin (top) and a silver pendant (bottom).
    }
    \label{fig:fasthyde}      
\end{figure*} 

\textbf{\subsection{Denoising Algorithm}}
\label{subsec:denoising}

In order to deal with bands severally corrupted by strong noise, a state-of-the-art HS image denoising method, FastHyDe, proposed by Zhuang \textit{et al.} in \cite{2018_Zhuang_FastHyDe} is adjusted and tested on THz-TDS images. HS denoising problem is formulated following (\ref{equ_obs_model_H}) with $\textbf{H} = \textbf{I}$, where $\textbf{I}$ represents the identity matrix.

Considering the high correlation between bands of THz-TDS images, same as in FastHyDe, the following assumption is considered: the spectral vectors $\textbf{x}_i$, for $i = 1,...,n$, live in a $p$-dimensional subspace $\mathcal{S}_p$, where $p \ll b$. 
Following this assumption and defining $\textbf{E} = [\textbf{e}_1,...,\textbf{e}_p] \in \mathbb{R}^{b \times p}$ as a basis for $\mathcal{S}_p$, $\textbf{X}$ can be written as 
\begin{equation}
\label{equ:x_ea}
\textbf{X} = \textbf{E} \textbf{A},
\end{equation}
where $\textbf{A} \in \mathbb{R}^{p \times n}$ corresponds to the representation coefficients of $\textbf{X}$ in $\mathcal{S}_p$. In FastHyDe, the rows of $\textbf{A}$ are called eigen-images. The above assumption on the high correlation between bands (i.e., the assumption that HS data follow a low-rank structure) is crucial, as matrix $\textbf{E}$ may be learned directly from $\textbf{Y}$ by subspace identification methods such as HySime \cite{2007_Nascimento_HySime}. Bands of HS images (including THz time-domain images) are self-similar in the spatial domain, i.e., they contain many similar patches at different locations and scales. The exploitation of self-similarity as a form of prior knowledge is the second main assumption of the FastHyDe method applied through the exploitation of a patch-based denoising step. The main steps of FastHyDe are shown in Algorithm 1.

\begin{algorithm}[h]
\small
\caption{FastHyDe}
\begin{algorithmic}
 \STATE  1) Learn the subspace $\textbf{E}$ from $\textbf{Y}$\;
 \STATE  2) Compute noisy eigen-images $\textbf{E}^T\textbf{Y}$\;
 \STATE  3) Denoise eigen-images by a state-of-the-art denoiser \cite{2007_Dabov_Image}\;
 \STATE  4) Compute an estimate of the clean HS image
 \end{algorithmic}
\label{alg1}
\end{algorithm}

The results of FastHyDe applied to all bands jointly of two real THz time-domain images are displayed in Figure \ref{fig:fasthyde}. Note that only selected bands are showed. The results are obtained by assuming Gaussian i.i.d. noise and the number of subspaces $p = 10$. The tested method is able to deal even with severe noise that corrupt bands corresponding to higher frequencies (e.g. frequencies higher than 4 THz). At lower frequencies, the output is comparable to the input, as the tested method only removes noise and not beam shape effects.

\textbf{\subsection{Joint denoising and deblurring Algorithm}}
\label{subsec:joint_processing}

The fast deblurring and denoising method \cite{2020_Ljubenovic_Joint}, tailored to THz time-domain images, three noise types (Gaussian i.i.d., Gaussian non-i.i.d., and Poisson), and band-dependent blur is further tested on 1 cent and a silver pendant. The method follows the same low-rank assumption from (\ref{equ:x_ea}) but instead of performing only denoising on subspace components (i.e., eigen-images) it additionally adds a non-blind deblurring step (i.e., deblurring with a known PSF). Note that deblurring in this case is performed only on subspace components and not on all bands separately (as in case of methods presented in Subsection \ref{subsec:2d_deblurring}A).  
Consequently, the joint deblurring and denoising method consists of steps presented in Algorithm 2.  

\begin{algorithm}[]
\small
\caption{Joint deblurring and denoising of THz time-domain images}\label{alg:alg1}
\begin{algorithmic}
\STATE 
\STATE {\textsc{Dimensionality reduction}}
\STATE \hspace{0.5cm} 1) Learn the subspace $\textbf{E}$ from $\textbf{Y}$
\STATE \hspace{0.5cm} 2) Compute noisy eigen-images $\textbf{E}^T\textbf{Y}$
\STATE {\textsc{Blur and noise removal}}
\STATE \hspace{0.5cm} 3) Analyse subspace domain
\STATE \hspace{0.5cm} 4) Create synthetic PSFs
\STATE \hspace{0.5cm} 5) Deblur eigen-images with an off-the-shelf non-blind deblurring method
\STATE \hspace{0.5cm} 6) Denoise eigen-images by a state-of-the-art denoiser from \cite{2007_Dabov_Image}
\STATE {\textsc{Reconstruction}}
\STATE \hspace{0.5cm} 7) Compute an estimate of the clean HS image.
\end{algorithmic}
\label{alg2}
\end{algorithm}

\begin{figure*}[!t]
    \centering
    \includegraphics[width=0.95\textwidth]{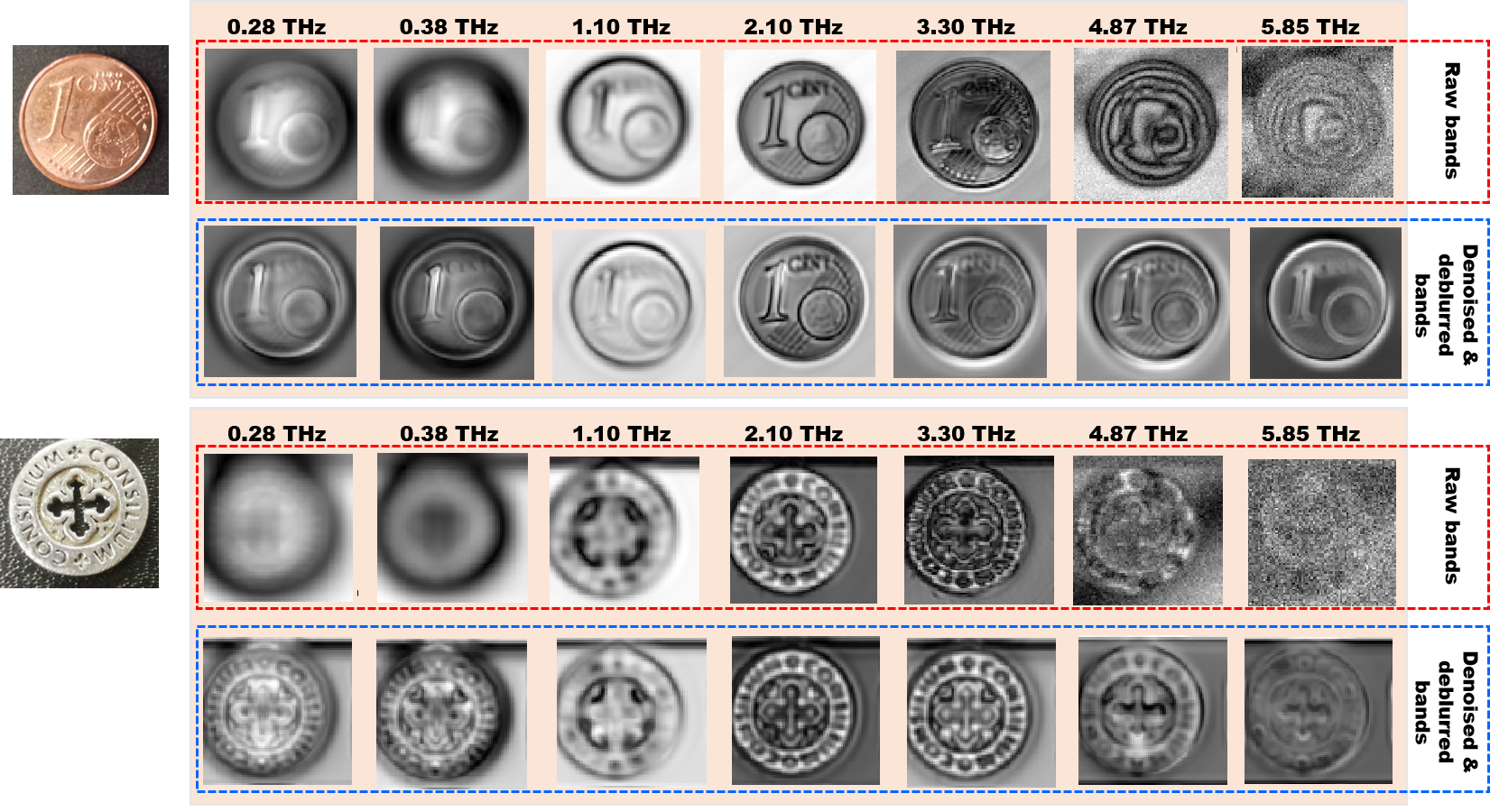}
    \caption{Results of the joint deblurring and denoising method for restoring amplitude images, as described in Algorithm 2. The raw results are compared with the images obtained after restoration for single THz frequency values. The method is applied on two metallic samples: 1 Euro cent coin (top) and one silver pendant (bottom).}
    \label{fig:joint_de}      
\end{figure*}

\begin{figure}
    \centering
    \includegraphics[width=0.35\textwidth]{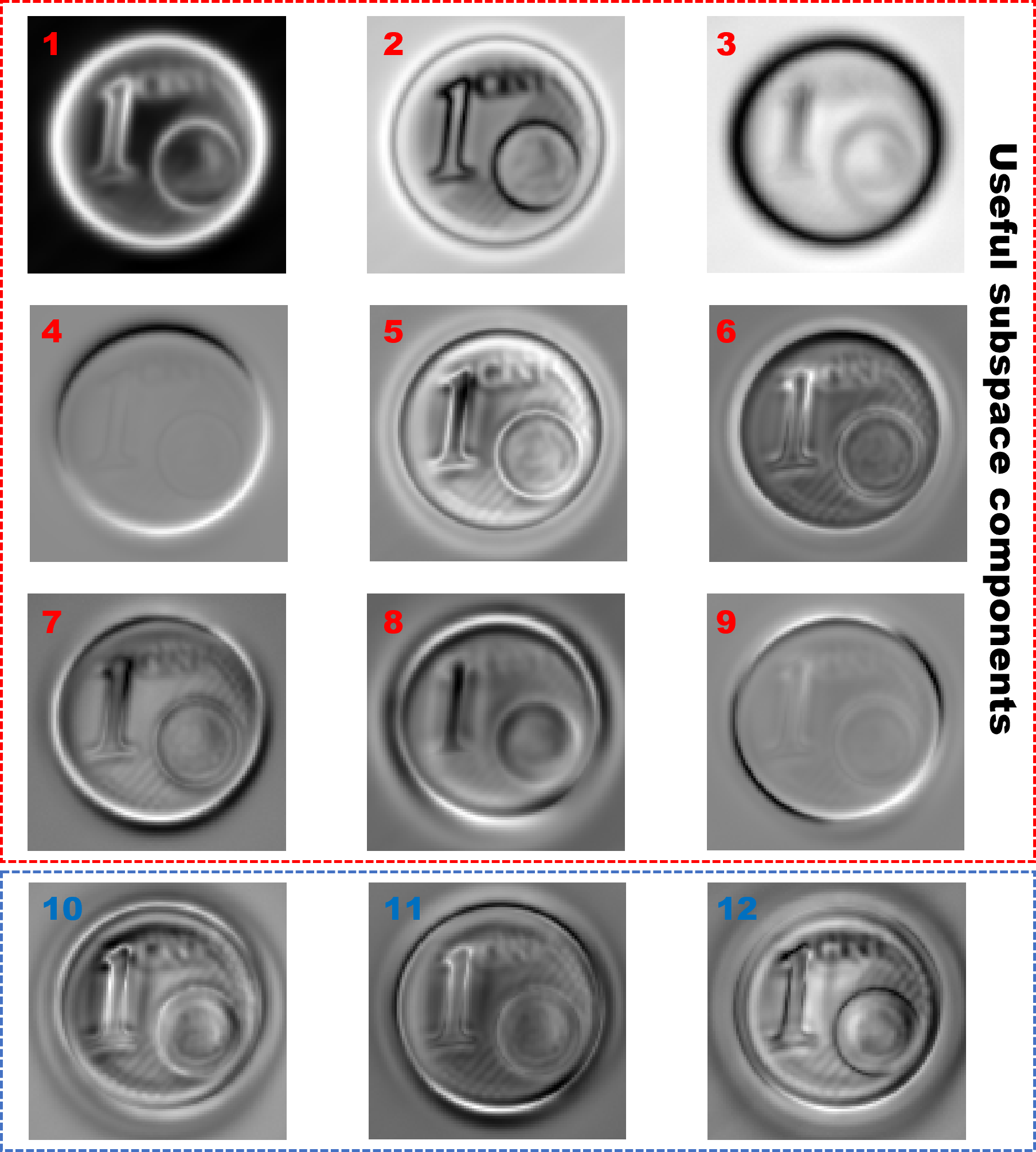}
    \caption{Subspace components of 1 Euro cent ($p = 12$): Subspace components from 1 to 3 cover the main edges; components from 5 to 7 correspond to textural features; components 4, 8, and 9 account for flat regions; component from 10 to 12 are showing effects of reflection.}
    \label{fig:eigen_im}
\end{figure}

The results obtained with one instance of the joint deblurring and denoising method with Gaussian i.i.d. noise assumption and Richardson-Lucy deblurring step (Step 5 in Algorithm 2) are shown in Figure \ref{fig:joint_de}. 
With respect to the transmission mode, here a new step is included covering detailed analysis of the subspace domain and corresponding eigen-images (Step 3 in Algorithm 2). 
In \cite{2020_Ljubenovic_Joint}, only samples with simple shapes and textures are tested thus, the number of subspaces can be fixed for all samples. 
Here, the number of subspaces is carefully chosen to include all significant features (e.g., main edges and texture) and exclude features corresponding to reflection effects (see Fig. \ref{fig:eigen_im}). 
Similarly, the values of the minimum beam radius utilized for PSF modelling is carefully set by analysing blur degradation of eigen-images, i.e., eigen-images corresponding to main edges (e.g., images 1, 2, and 3 in Fig. \ref{fig:eigen_im}) are deblurred by using bigger PSFs relative to the eigen-images corresponding to textural features (e.g., images 5, 6, and 7 in Fig. \ref{fig:eigen_im}). The subspace components corresponding to reflection effects (images 10, 11, and 12) are discarded during the analysis and thus, the number of subspace components for the 1 cent sample is set to $p = 9$. The similar analysis of subspace components is repeated for each sample. 

As the tested joint deblurring and denoising method is based on FastHyDe, the noise from bands corresponding to higher frequencies is removed equally successful as previously. Additionally, beam shape effects removal is visible in bands corresponding to lower frequencies (i.e., frequencies less than 1 THz). 

\begin{figure}[!t]
    \centering
    \includegraphics[width=0.415\textwidth]{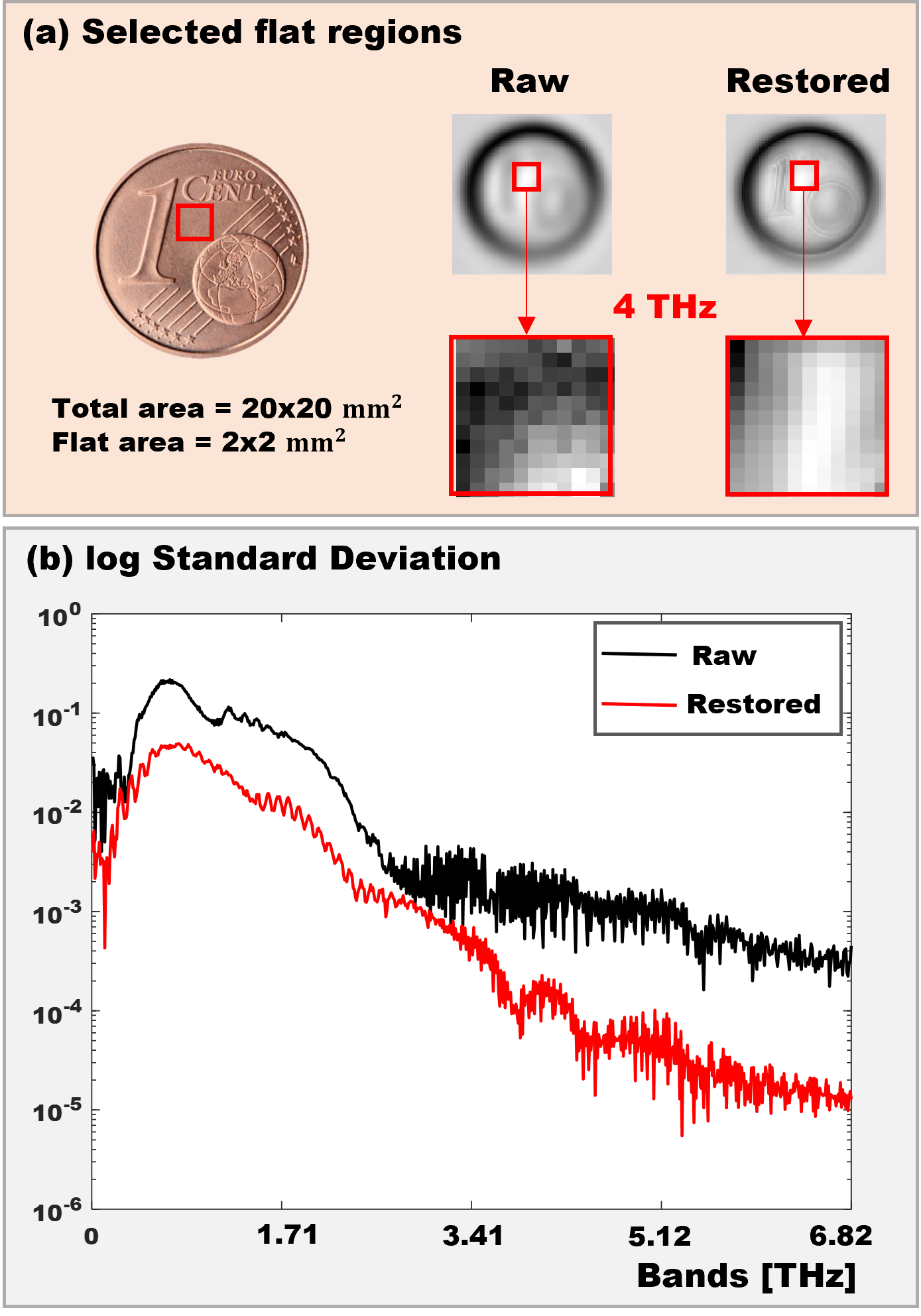}
    \caption{Performance of the joint deblurring and denoising method measured through standard deviation. (a) A selected flat region from raw and restored bands for calculated standard deviation; (b) log standard deviation as a function of different frequencies.}
    \label{fig:std}      
\end{figure}

\begin{figure}[!t]
    \centering
    \includegraphics[width=0.48\textwidth]{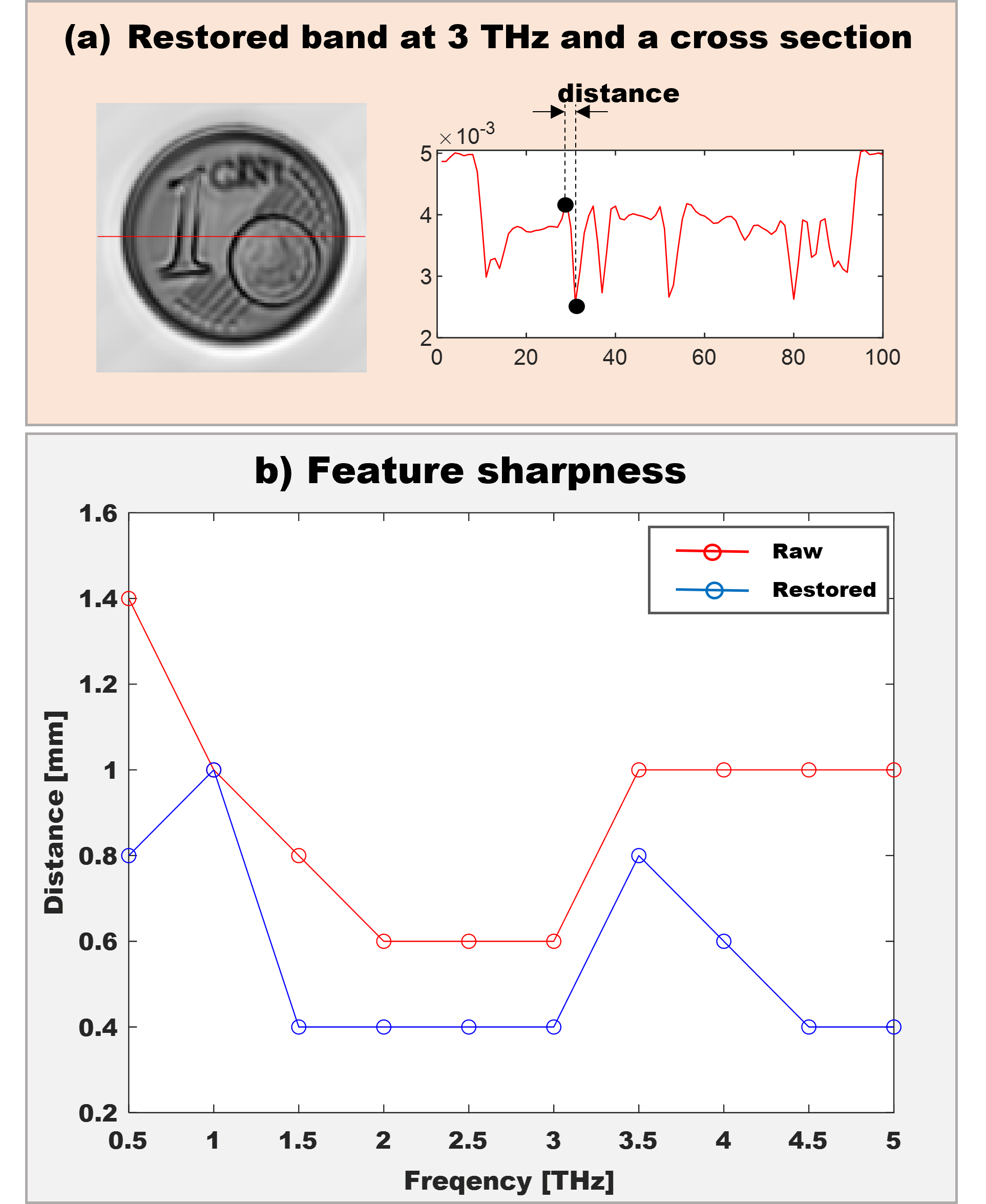}
    \caption{Performance of the joint deblurring and denoising method measured through a sharpness of a feature (the number one on the coin). (a) Restored band at 3 THz and its cross section; (b) Feature sharpness estimated as a distance between a high and a low peak.}
    \label{fig:resolution}      
\end{figure}

The performance of the joint deblurring and denoising method applied to images acquired in reflection geometry was measured by comparing standard deviation of an even region (Fig. \ref{fig:std}) and feature sharpness (Fig. \ref{fig:resolution}) of raw and restored bands. The standard deviation was estimated over a $2 \times 2$ mm homogeneous region selected from the 1 euro cent coin. For each pixel, the variance of reflected reference intensity (raw data) and reconstructed intensity (estimated data) is computed and compared. The improvement resulting from the restoration procedure is visible over all bands, increasing with the increase of frequency. 
Similarly, a feature sharpness (in mm) is estimated by measuring a distance between a high and a low peak that represent a border of the feature. Decrease in distance indicate the steepest feature border and thus, a sharper feature. The distance measure is highly affected by the presence of noise, especially visible on the higher frequency bands.


Both visual and the numerical results (i.e., standard deviation and feature sharpness) respectively, demonstrate that the joint deblurring and denoising approach achieves significantly improved THz HS images of homogeneous materials without introducing additional distortions.
The reconstruction approach was able to restore both lower (e.g., below 1 THz) and higher (e.g., above 3.5 THz) spectral components, making them more suitable for further analysis.
Successful restoration of bands corresponding to the broad THz spectral range drastically increase the amount of information on a sample, enabling both spatial and spectral domains to be fully utilized in different applications such as cultural heritage, biomedical imaging, quality inspection, etc. 
\\

\section{Application of the method to the study of an ancient coin}
\label{sec:thz_in_ch_res}

\begin{figure}[!t]
    \centering
    \includegraphics[width=0.45\textwidth]{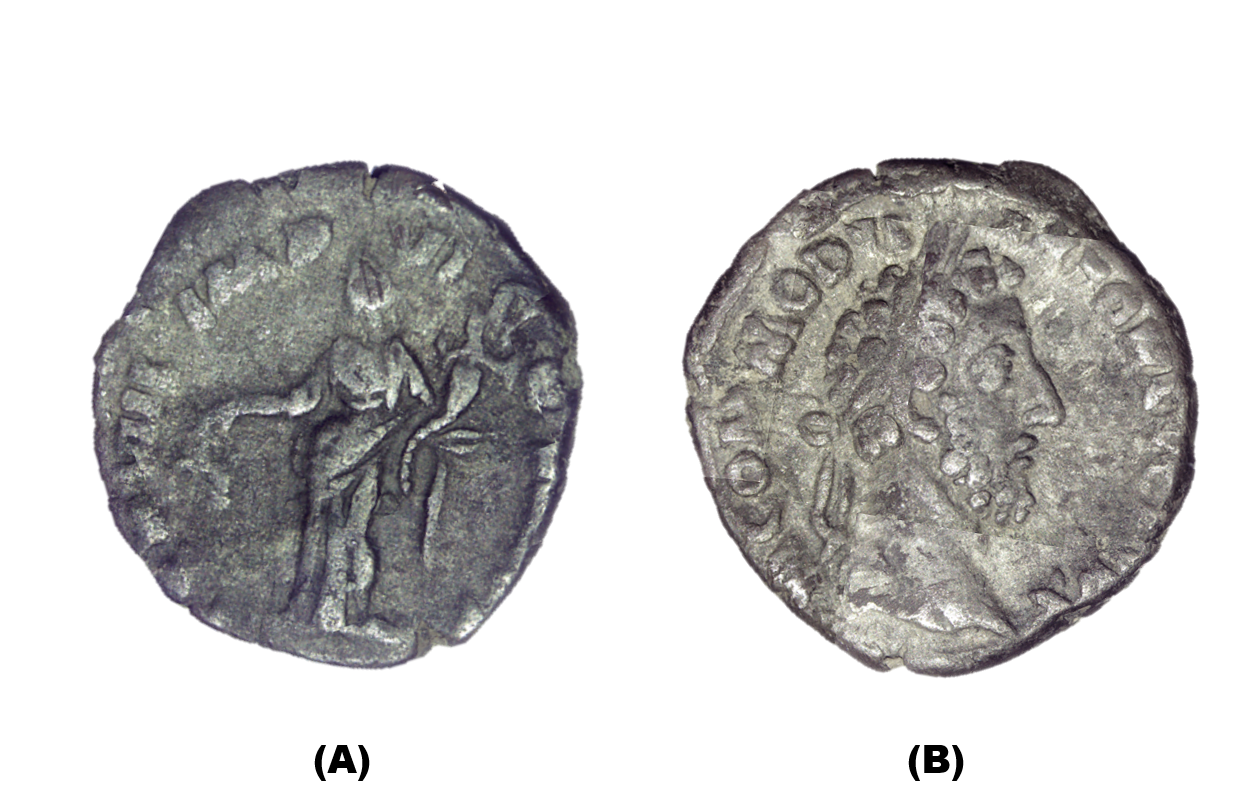}
    \caption{Stereoscopic images of the archaeological silver coin. On the front side the figure of a goddess is inlaid \textbf{(A)}, while on the back side the one of a Roman emperor portrait \textbf{(B)}.}
    \label{fig:roman_coin}       
\end{figure}

Amelioration results obtained on THz-TDS images of the two objects show the potential of the image restoration procedures in reflection geometry. In particular, they enable to retrieve clean images at very high (above 3.5 THz) and very low (below 1 THz) frequencies, otherwise not accessible due to heavy noise and blur effects. This represents a highly desirable result, especially when dealing with material characterisation. 
To test the capacity of the algorithm on more complex objects (in terms of shape and material composition), the archaeological coin showed in Fig. \ref{fig:roman_coin} is utilized.

The ancient coin used in this study has worn-out reliefs, a surface altered by the ageing process, and is encrusted with soil residue. The opacity to the THz radiation of this type of object requires the acquisition of THz images in reflection geometry, but the presence of alteration layers on the surface can induce multiple reflections and can heavily affect the amplitude signal. The analysis was undertaken in the same way as for the contemporary objects illustrated in the previous sections and both sides of the Roman coin were included. 

\begin{figure*}[!t]
    \centering
    \includegraphics[width=1\textwidth]{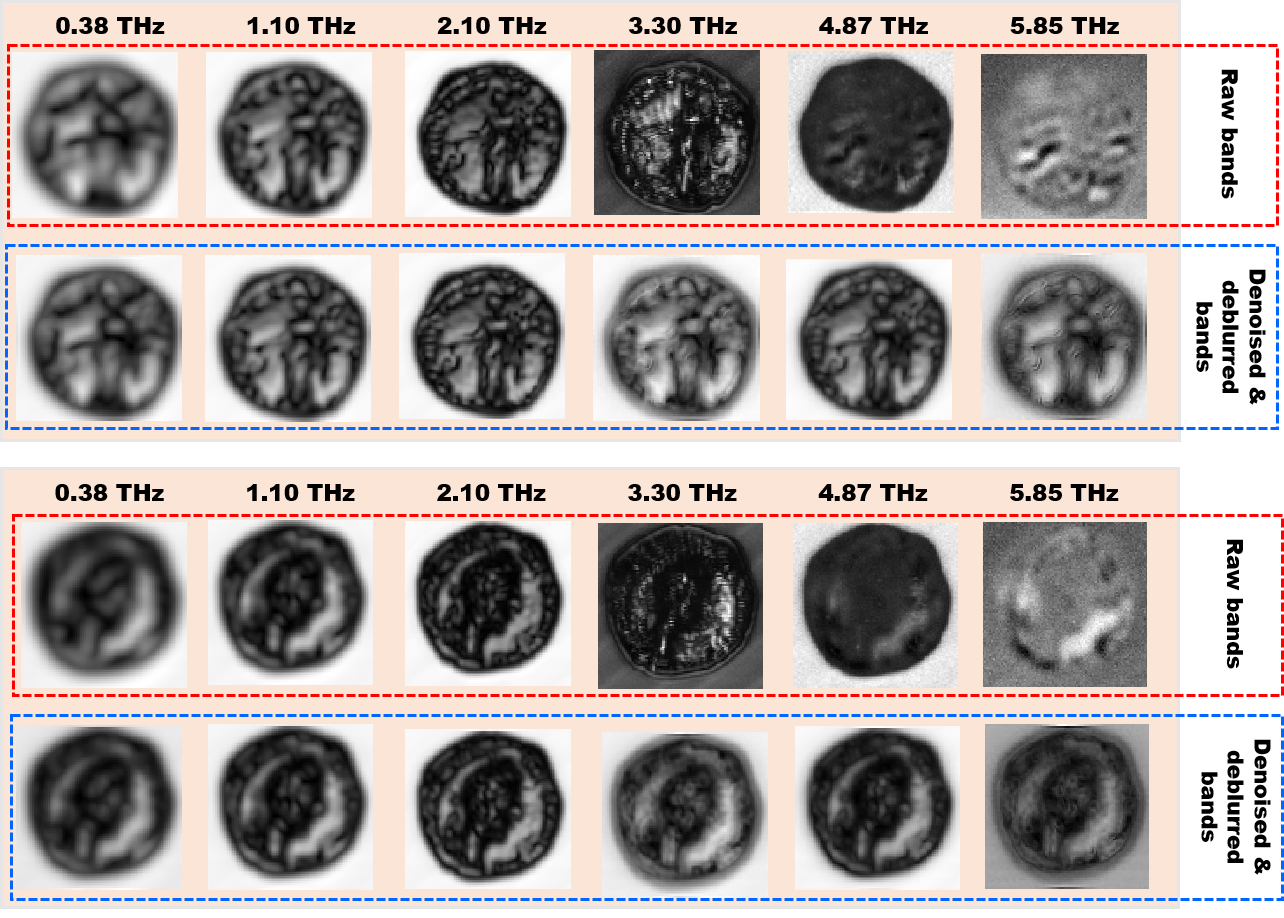}
    \caption{Results of the joint denoising and deblurring method on the front and back faces of the archaeological Roman coin. 
    The results are showed as integrated amplitude images over the frequency range which starts from the frequencies presented above the images + 0.4 THz.}
    \label{fig:roman_coin_res}       
\end{figure*}

\begin{figure*}[!t]
    \centering
    \includegraphics[width=0.9\textwidth]{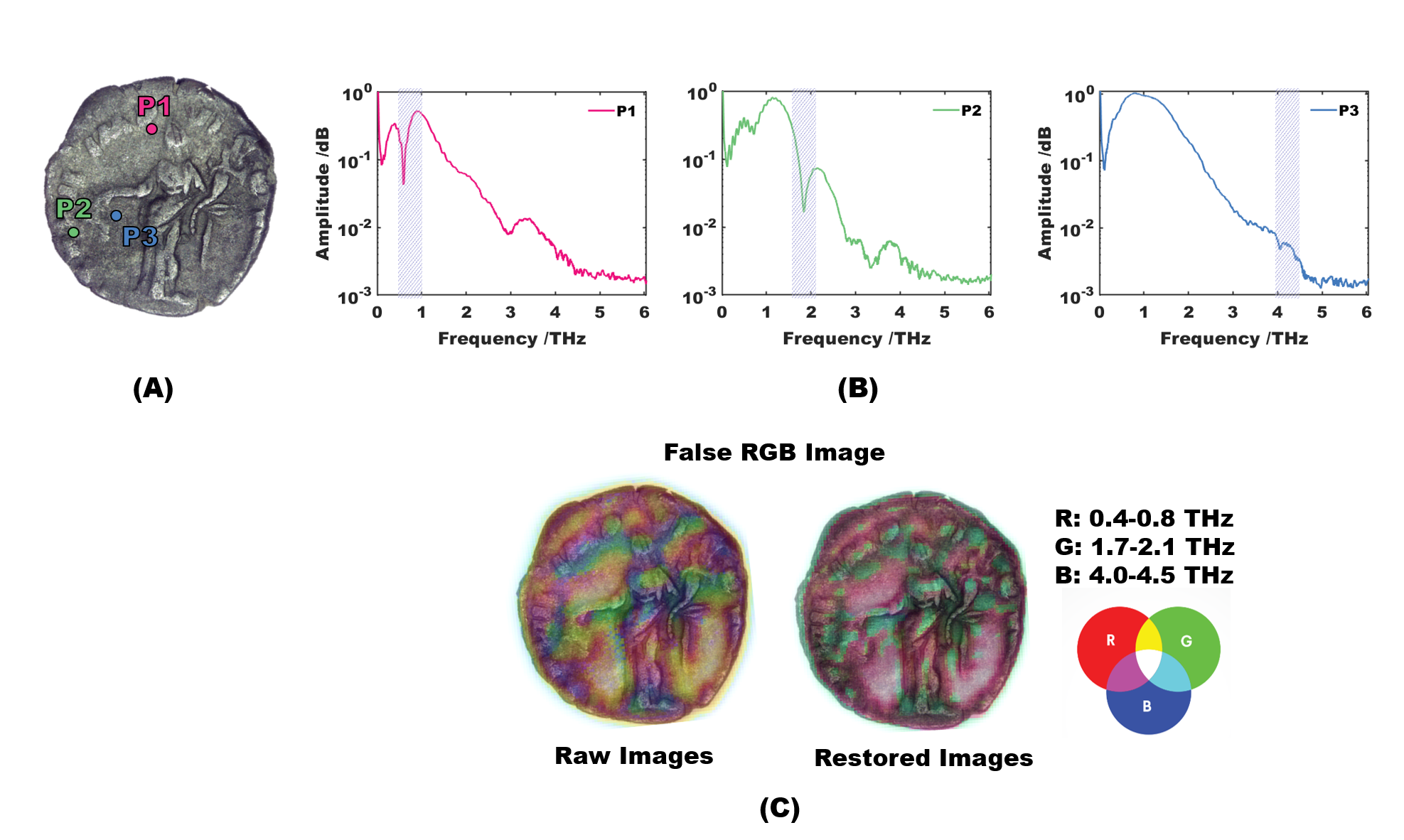}
    \caption{False RGB image of raw data and after image restoration, constructed from three integrated amplitude images. The red channel (R) corresponds to 0.4-0.8 THz, which is a highly blurry frequency region, the green channel (G) is integrated amplitude between 1.7-2.1 THz, which shows a good quality result and high dynamic range, and lastly the blue channel (B) corresponds to 4.5-5.5 THz, more affected by noise. The false RGB image after restoration delineate better the distribution of the amplitude intensity in the three frequency regions.}
    \label{fig:roman_coin_rgb}       
\end{figure*}

The image restoration approach used in this experiment is the one presented in Section \ref{subsec:joint_processing} with assumption of Gaussian i.i.d. noise.  The number of subspaces and  $w_0$ corresponding to PSFs are chosen by analysing subspace components as explained in Subsection \ref{subsec:joint_processing}. As for the other samples previously investigated, the conventional Richardson-Lucy algorithm is used as a deblurring step and the state-of-the-art patch-based method from \cite{2007_Dabov_Image} as a denoising step. The integrated amplitude values for six frequency ranges of raw and restored THz HS images are presented in Figure \ref{fig:roman_coin_res}: the high variety between bands (from blurrier to noisier ones) demonstrate the need for a robust approach when dealing with cultural heritage objects. 
Compared to contemporary samples, that are made by known and homogeneous composition, Roman coins are made of bronze or silver alloy, which developed metallic oxide on the surface. The spectral signatures of these compounds can be highly challenging to determine in the THz domain. Thus, the results in Figure \ref{fig:roman_coin_res} are represented as an integrated amplitude over a range of frequencies (starting from the frequencies presented in the figure plus 0.4 THz).
The restoration procedure results in sharpening the bands corresponding to lower frequencies (e.g., 0.38-0.78 THz) and fully removing severe noise from bands corresponding to higher frequencies (e.g., 5.85-6.25 THz), thus expanding the usable frequency range for further analysis. 

The highly modular and robust approach for joint deblurring and denoising utilized in this work for the analysis of objects with irregular shape and complex surface texture provides highly promising results. The modularity of the approach, reflected in its step-like design, ensures a better evaluation of the performance of each separate step and therefore an easier tuning of input parameters. 
The patch self-similarity assumption makes the approach more robust to irregular shapes of the sample, representative of the shapes normally found in cultural heritage items: the approach is focused to image patches (small image parts) rather than the whole image, thus separately capturing wider complexity of shapes and textures.
 
The final result is represented by a false-color image, where three amplitude images at very low frequency (0.4-0.8 THz), in the medium region (1.7-2.1 THz) and at very high frequency (4.5-5.5 THz), are merged, as illustrated in Fig. \ref{fig:roman_coin_rgb}. Outputting a false-color image is quite common in HS image analysis, however in this case THz-TDS raw results could in principle induce to wrong conclusions because the presence of blurring and noising effects can alter the features of the objects. After the image restoration, the resolution of the images become similar and the distributional map more reliable. This work does not include comparison of the results to a reference (e.g., a spectra from a spectral signature library), and thus the features presented could be topographical.


\section{Conclusion}
The paper demonstrated the positive outcomes achieved using a computational procedure for THz-TDS imaging to restore amplitude signals in reflection geometry and to tailor it to deal with irregular surfaces and uneven reliefs, such as the ones of the archaeological item illustrated in this work.  

Utilizing image restoration approaches that are able to handle strong noise is crucial when applying THz time-domain imaging to objects that have distinguishable spectral features in the middle and high frequencies (i.e., frequencies corrupted by such noise).
The importance of applying image processing strategies to HS images is most evident for applications that relies on the possibility of extension of the available frequency range using far-infrared frequencies (spanning from few hundreds of GHz to 10 THz) enabling a more  comprehensive analysis.
This study broke the 3 THz limitation, defined by the previous studies \cite{Manceau2008}, and showed reliable image reconstruction over more than 4 octaves, from 0.28 THz and reaching up to 5.85 THz.



Further work in this field will have to entail the development of a novel THz super-resolution approach to overcome the resolution limitation partially imposed by a step size of the instrument during acquisition and significantly increase the size and number of samples suitable for analysis.  

\section*{Acknowledgment}
Images of the ancient Roman coin are courtesy of the Soprintendenza Archeologia, belle arti e paesaggio del Friuli Venezia Giulia  (prot. SABAP N.0004987-P, 30/12/20), which the authors wish to thank for the kind support and availability.


 
\bibliographystyle{IEEEtran}
\bibliography{IEEEabrv,thz_ref}

\vfill

\end{document}